\documentstyle[twocolumn,aps]{revtex}

\input epsf

\begin{document}
\title{Cooper Pairing in Ultracold $^{40}$K Using Feshbach Resonances}
\author{John L. Bohn}
\address{JILA, National Institute of Standards and Technology and
University of Colorado, Boulder, CO 80309-0440}
\date{\today}
\maketitle

\begin{abstract}
We point out that the fermionic isotope $^{40}$K is
a likely candidate for the formation of Cooper pairs in
an ultracold atomic gas. Specifically, in an optical 
trap that simultaneously traps the spin states
$|9/2,-9/2 \rangle$ and $|9/2,-7/2 \rangle$, there exists
a broad magnetic field Feshbach resonance at 
$B=196_{-21}^{+9}$ that can provide the
required strong attractive interaction between atoms.
An additional resonance, at $B=191_{10}^{+5}$ gauss,
could generate p-wave pairing between identical
$|9/2,-7/2 \rangle$ atoms.
A Cooper-paired degenerate Fermi gas could thus
be constructed with existing ultracold atom technology.
\end{abstract}


\narrowtext

Recently an ultracold gas of fermionic $^{40}$K atoms
was cooled to the quantum degenerate regime \cite{DeMarco1}.
This achievement  opens a new chapter in the story of
ultracold matter, complementary to the 
Bose-Einstein condensation work that has been going on for
over four years now.  The degenerate Fermi gas (DFG) is
expected to exhibit novel behavior in its thermodynamics 
\cite{Bruum}, collision dynamics \cite{Ferrari}, and scattering
of light \cite{DeMarco0,Busch}.
Perhaps the most intriguing prospect for the DFG is the
potential to observe a pairing of the fermions, leading
to a derived superfluid state, analogous to the 
Cooper pairing of electrons in a superconductor
\cite{Modawi,Kagan,Stoof,Ho,Zhang}.

To make such a pairing work requires an effective attraction
between colliding atom pairs in the  gas.  For bosons, 
an attractive interaction corresponds to a negative value 
of the s-wave scattering length.  For fermions, however,
the Pauli exclusion principle prohibits s-wave scattering
of atoms in identical spin states.   This leaves
only p-wave collisions as a pairing mechanism, but the
resulting interactions are energetically suppressed and 
are generally considered to give experimentally unattainable 
pairing transition temperatures \cite{Leggett}.  On the other hand,
a recent proposal has suggested that p-wave interactions
may be enhanced by the application of very large dc electric
fields, which could be generated by powerful CO$_2$ lasers \cite{LiYou}.

A second possibility would be to use two different
spin states of a fermionic atom,  thus restoring s-wave
collisions as a pairing mechanism.  In this context $^6$Li
appears to be an attractive candidate \cite{Houbiers}, 
since it possesses a large negative s-wave scattering length 
$a_s = -2160$ $a_0$, in units of the Bohr radius $a_0$  \cite{Li6}.
[In this paper we distinguish s-wave and p-wave scattering lengths
with the subscripts ``$s$'' and ``$p$''.  To
avoid confusion with standard notations, we indicate 
singlet and triplet explicitly in superscripts, as in Eq. (3), below.]
In this case the critical 
temperature for Cooper pairing is approximately \cite{Houbiers}
\begin{equation}
T_c \sim {E_F \over k_B}
\exp \left( -{\pi \over 2 k_F |a_s|} \right),
\end{equation}
where $E_F$ and $k_F$ are the Fermi energy and momentum,
respectively.  As pointed out in Ref. \cite{Houbiers}, for experimentally 
realizable Fermi energies $E_F/k_B \sim 600$ nK 
there would result $T_c \sim 15$ nK for $^6$Li.  Thus any
alkali atom with a similarly large, negative scattering length
should be a viable candidate for Cooper pairing.

The purpose of this paper is to consider the prospects for Cooper pairing in
$^{40}$K, in both s-waves and p-waves, based on a magnetic-field 
Feshbach resonance that can be used to tune its scattering length.  
This atom has been trapped and cooled in several labs
\cite{DeMarco1,Cataliotti}.
The ability to tune scattering lengths resonantly using magnetic
fields \cite{Stwalley,Eite} is now a proven technology.  
To date, this resonant tuning has been observed in Na \cite{Ketterle},
Rb \cite{Heinzen,Roberts}, and Cs \cite{ChuCs}.  That these
resonances are in fact useful tools for manipulation of
ultracold gases has been amply demonstrated recently
in an experiment that used them to Bose-condense the otherwise
uncondensible $^{85}$Rb isotope \cite{RobertsBEC}.

Although the ``required'' scattering length to ensure formation of
Cooper pairs will depend strongly on experimental
circumstances, we can estimate a reasonable set of parameters
using the guidelines laid out in Ref. \cite{Houbiers}.  A first
requirement is that the resulting Cooper-paired state be
mechanically stable, which for a two-component gas with
number densities $n_1$ and $n_2$ requires \cite{Houbiers}
\begin{equation}
n_1 n_2 a_s^6 \leq \left( {\pi \over 48} \right)^2.
\end{equation}
When the scattering length violates this condition, two
kinds of instability may occur: if $a_s < 0$, at least one
component collapses into a dense, probably solid state; whereas
if $a_s > 0$, the two components will phase-separate \cite{Houbiers}.
Note that with a tunable $a_s$ these instabilities can be probed
experimentally in $^{40}$K.

If we assume equal densities of $n_1 = n_2 = 10^{14}$ cm$^{-1}$, then
Eq. (2) imposes the restriction $|a_s| < 1700$ $a_0$.  Conservatively,
we will adopt a target value of $a_s = -1000$ $a_0$ in the following.
In this case, for a Fermi temperature of $T_F \sim 600$ nK 
(compare Ref. \cite{DeMarco1}) we would find a Cooper pairing 
temperature of $T_c \sim 25$ nK in $^{40}$K.  Moreover, we are
interested in the stability of this $T_c$ against variations
in the magnetic field strength.  Let us require that $T_c$
remain constant to within a small fraction, say $10 \%$.
Eq. (1) then tells us that we must maintain $a_s$ constant 
to within $\sim 3 \%$.  We will see below that this criterion 
should be relatively easy to meet for the resonance described.

To compute Feshbach resonances in $^{40}$K,
we employ the standard close-coupled Hamiltonian for 
ultracold alkali-atom scattering \cite{Burkethesis}.
As usual, meaningful results can be obtained from this
Hamiltonian only if it is fine-tuned with the help
of experimental data.  In this case we will employ
the constraints imposed by a recent analysis of photoassociation
spectroscopy of the $0_g^-$ state of $^{39}$K$_2$ 
\cite{BurkePA}.  This analysis reveals a $^{39}$K
triplet scattering length (in $a_0$) of
\begin{equation}
a_s^{\rm triplet}(39) = -17 - 0.045 (C_6-{\bar C}_6) \pm 25,
\end{equation}
with ${\bar C_6} = 3800$ atomic units \cite{Marinescu}.
This parametrization allows for an uncertainty in the $C_6$
coefficient that determines the long-range van der Waals
attraction between the atoms.  The experiment itself provides
no direct information on the value of $C_6$.
The result in Eq. (3) is consistent with a complementary analysis
of the $1_u$ state of $^{39}$K, which gives
$-60$ $a_0$  $ < a_s^{\rm triplet}(39) <$ $15$ $a_0$ \cite{NISTPA}.

Rescaling by the appropriate reduced mass, Eq. (3)
implies for $^{40}$K a nominal triplet scattering length
of $a_s^{\rm triplet}(40) = 176$ $a_0$.
This result is consistent with the values obtained in
a direct collisional measurement in $^{40}$K \cite{DeMarco2}.
Finally, we take the singlet scattering
length to be $a_s^{\rm singlet}(40) = 105$ $a_0$  \cite{BurkePA,BohnK}.  
This value is fairly well constrained by the existing data; moreover, the 
results of this paper depend only weakly on its exact value.

There remains the issue of the value of $C_6$ to employ
in the calculations.  The results of Marinescu {\it et al.}
cover the fairly broad range $C_6 = 3800 \pm 200$ atomic units
\cite{Marinescu}.
The accuracy of this result is limited by uncertainties
in the atomic data used in the calculation.  By contrast,
a new high-precision calculation by Derevianko {\it et al.} 
predicts a much narrower range of $C_6 = 3987 \pm 15$
\cite{Babb}.  This improvement is largely due to 
Derevianko {\it et al.}'s accurate calculation of atomic
structure, which freed them from experimental uncertainties.
Their track record is impressive: for Na \cite{Na} 
and Rb \cite{Roberts}, their predictions are within experimental
uncertainty of inferred values of $C_6$.  This result lends
credence to their value of $C_6$ for potassium, which we will adopt here.
In this case the largest uncertainty in potassium scattering
lengths arise from the $\pm 25$ in Eq. (3), rather than
from $C_6$.  Taking this uncertainty into account, and rescaling
the mass, the $^{40}$K triplet s-wave scattering length 
is given by $a_s^{\rm triplet}(40) = 176_{-27}^{+77}$ $a_0$.

Perhaps the most appealing candidate spin states in which to seek
a Feshbach resonance would be the magnetically trappable states
$|fm \rangle=$ $|9/2,9/2 \rangle$ and $|9/2, 7/2 \rangle$,
which are already trapped in the JILA experiment \cite{DeMarco1}.
However, as reported in Ref. \cite{BohnK}, no such resonance exists.
There may be resonances for nearby spin states, but these
should be very narrow ($\Delta B \ll 1$ gauss) and probably
not useful for Cooper pairing.

There is, however, a broad resonance in collisions between
the states $|9/2,-9/2 \rangle$ and $|9/2, -7/2 \rangle$,
as illustrated in Fig. 1.  This resonance, lying between
175 and 205 gauss, is easily accessible experimentally.
Moreover its broad width implies that the scattering length 
can be tuned quite accurately. The inset to Fig. 1 focuses on
the region near $a_s = -1000$ $a_0$. To maintain this value of
the scattering length  to within $3 \%$ (i.e., to
maintain $T_c$ constant to within $10 \%$, as discussed
above) would require
holding $B$ steady to within $\sim 0.1$ gauss. 
Since the two states are strong-field seekers, they cannot
be trapped in the usual magnetic traps that have traditionally
been used for BEC studies.  Nevertheless, the two states could 
be held in an optical trap.  These traps have recently
attained great stability, with lifetimes exceeding 
300 seconds \cite{Thomas}.  Note also that an optical trap
ensures that the magnetic field can be made uniform
across the entire trap, so that all atoms would
experience the same pairing interaction.  Evaporative cooling
may be possible is these traps, as well
\cite{Chu,Corwin}.

These particular spin states are also appealing in terms 
of their stability against collisional losses.
At the ultralow temperatures of interest here, p-wave
collisions are strongly suppressed, meaning that there
are virtually no losses due to collisions between atoms in the same
spin state.  Inelastic collisions that produce $|9/2, -5/2 \rangle$
states are also energetically forbidden, since the energy of this
state lies 2.3 mK higher in energy than the $|9/2, -7/2 \rangle$
state at the magnetic fields considered.  There would then remain
only the collision process
\begin{equation}
|9/2,-9/2 \rangle + |9/2,-7/2 \rangle
\rightarrow |9/2,-9/2 \rangle + |9/2, -9/2 \rangle.
\end{equation}
This collision cannot occur in a spin-exchange process, which must conserve
the sum $m_a + m_b =-8$ of the magnetic quantum numbers. 
Nor can it proceed by the spin-spin dipolar interaction \cite{spinspin}.
This is because an incident s-wave can only couple to a
d-wave final state in this processes, but d-waves are forbidden for 
identical final spin states.  Thus the mixture we envision is 
virtually immune to two-body loss processes.

This leaves us with the possibility for three-body
loss processes, where two bodies recombine into a molecule
with the other carrying away the binding energy.  These
processes can generally contribute to heating, trap loss,
or contamination with unwanted molecular states.  They have
been observed to exert a strong influence on Bose-Einstein
condensates, especially near Feshbach resonances \cite{Ketterle}.  
The age of quantitative calculation of three-body recombination
has just begun \cite{Esry}.  Nevertheless, we can argue that
these losses, too, are suppressed in this system.  Roughly this is because
any three-body collision in a two-component Fermi gas must involve 
two identical atoms.  Again invoking the exclusion
principle, these atoms must have a nonzero relative angular momentum,
which effectively keeps them apart, suppressing the
collision. Following the more careful hyperspherical treatment of
the type in Ref. \cite{Esry}, this would lead to a threshold law
where the three-body recombination rate vanishes at low $E$
as $E^{1/2}$, in contrast to the $E$-independent rate expected
for bosons.

Finally, we return to the subject of possible p-wave Cooper
pairing, similar to that envisioned in Ref. \cite{LiYou}.
Non-s-wave pairing is already known in superconductors
and in superfluid $^3$He.  However, the ability to produce
this pairing in a dilute, weakly interacting atomic gas,
and moreover to control the strength of coupling, would
enable detailed experimental and theoretical study,
as has already been the case for dilute Bose condensates.
In this case the pairing temperature, analogous to Eq. (1),
is given by
\begin{equation}
T_c \sim { E_F \over k_B} \exp \left(- {\pi \over
2(k_F |a_p|)^3 } \right).
\end{equation}
Here $a_p$ stands for the ``p-wave scattering length,''
defined by 
\begin{equation}
a_p^3 = \lim_{k \rightarrow 0} - {\delta_p(k) \over k^3} ,
\end{equation}
where $\delta_p (k)$ is the p-wave scattering phase shift
and $k$ is the wave number.
The cubic dependence on $a_p$ of the exponential in (5) places
more severe restrictions on $a_p$ than in the s-wave case.
For example, for $T_F = 600$ nK, setting $a_p = -1000$ $a_0$
in (5) would yield a critical temperature of only $T_c = 0.002$ nK,
whereas for $a_p = -1500$ we would get $T_c = 14$ nK.
In the latter case, if we again require that $T_c$ be
constant to within $10 \%$, we find that $a_p$ must be
constant to within $1 \%$.

For  $^{40}$K the naturally occurring value of the triplet
p-wave scattering length is $a_p^{\rm triplet}(40) = -100$ $a_0$, 
which is far too small to be of use.  Fortunately, there 
are Feshbach resonances in this case, too.
Generally speaking, these resonances lie
at approximately the same values of magnetic field as the s-wave
resonances, since each s-wave bound state that can resonate
is accompanied by a p-wave bound state at a nearby energy. 
For example, in p-wave collisions of $|9/2,7/2 \rangle$ and
$|9/2,5/2 \rangle$, there are extremely narrow resonances, as for
the s-wave case. 

We therefore again seek resonances in  states
with negative values of $m_f$.  In particular, for collisions
in a gas of pure $|9/2,-7/2 \rangle$ atoms,  we find
a fairly broad resonance at a position of $B = 191_{-10}^{+5}$ gauss, 
as illustrated in Fig. 2.  This resonance has nearly the same
shape as the familiar s-wave resonances, but with an additional
inflection when $a_p \sim 0$, arising from the cube-root dependence
of $a_p$ on $\delta_p$.  This resonance is also somewhat
narrower in magnetic field than the s-wave resonance
reported above.  In this case holding $a_p -1500$ $a_0$ constant to
within one percent requires holding the magnetic field
constant to within perhaps 0.001 gauss.  

In a given experiment the desired values of scattering lengths
may differ from the sample values we have
considered.  In this case, it is useful to present approximate
fitting formulas for the resonances, computed for the 
nominal interaction potentials.  For s-waves, this fit is
\begin{equation}
a_s \approx 164 - {1260 \over (B - 196.2)}.
\end{equation}
The p-wave fit, for $a_p<0$ and very near resonance, is
\begin{equation}
a_p \approx -600 - {21 \over (B - 191.02)}.
\end{equation}
In each case, the scattering length is in $a_0$ and
the field $B$ is in gauss.

In conclusion, these magnetic-field Feshbach resonances make
possible a variable Cooper-pairing interaction in ultracold
$^{40}$K gases.  Such interactions will enable detailed studies
of s- and p-wave superfluid states, including their instabilities,
at a level not possible before.  Significantly, to implement 
these resonances requires no technology beyond what is currently 
available. 

I thank D. Jin and C. Greene for useful discussions.  This work
was supported by the National Science Foundation.

\epsfxsize = 3in
\epsfbox{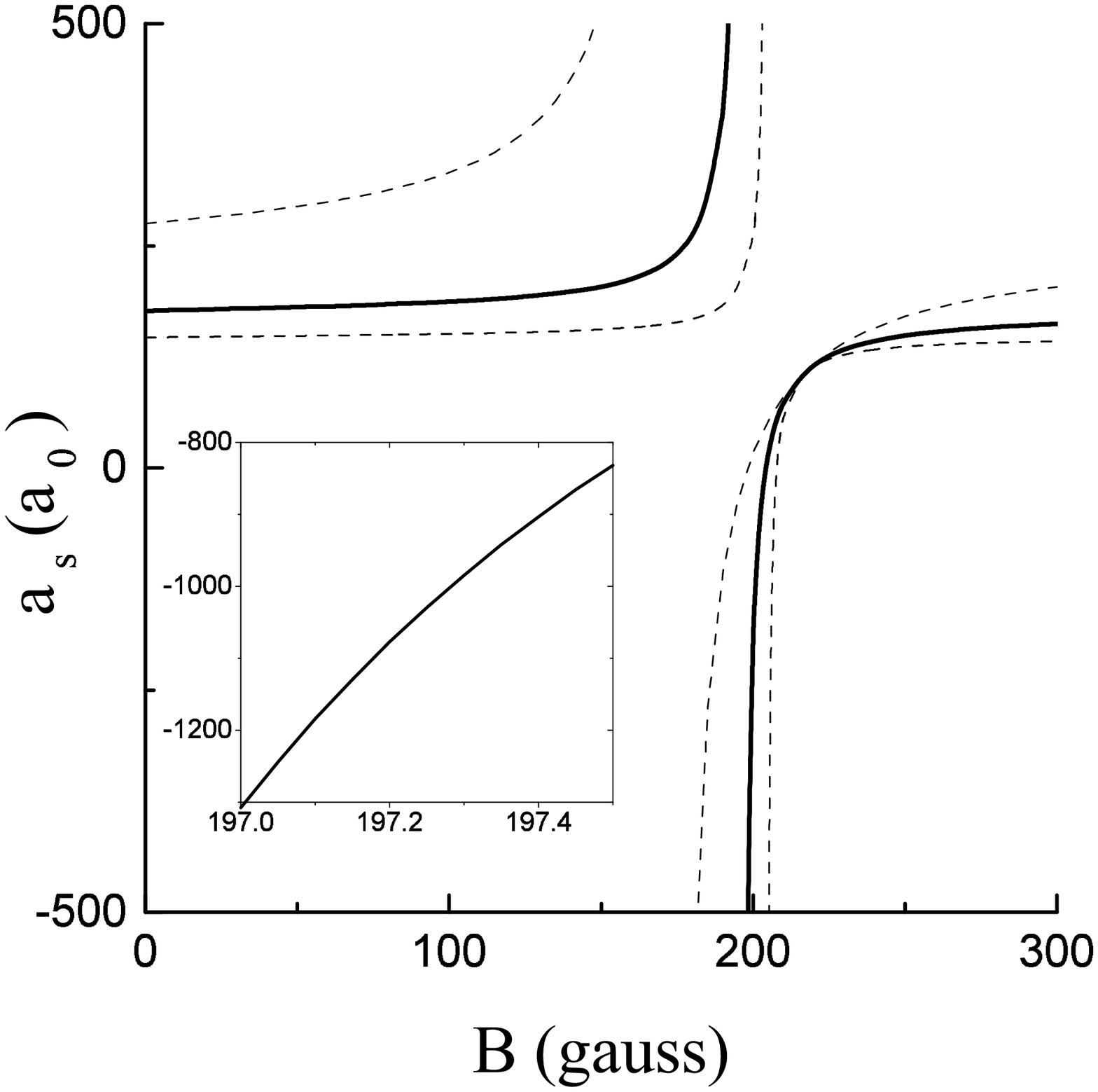}
\begin{figure}
\caption{Variation with magnetic field of the s-wave scattering length $a_s$
for $|9/2,-9/2 \rangle + $ $|9/2,-7/2 \rangle$ collisions of $^{40}$K.
The heavy line shows the nominal case, where $a_s^{\rm triplet}(40) = 176$ $a_0$,
while the dotted lines indicate the uncertainty of the resonance's
position through the uncertainty in $a_s^{\rm triplet}(40)$.  The inset shows
the nominal case in the vicinity of $a_s = -1000$ $a_0$.}
\end{figure}

\epsfxsize = 3in
\epsfbox{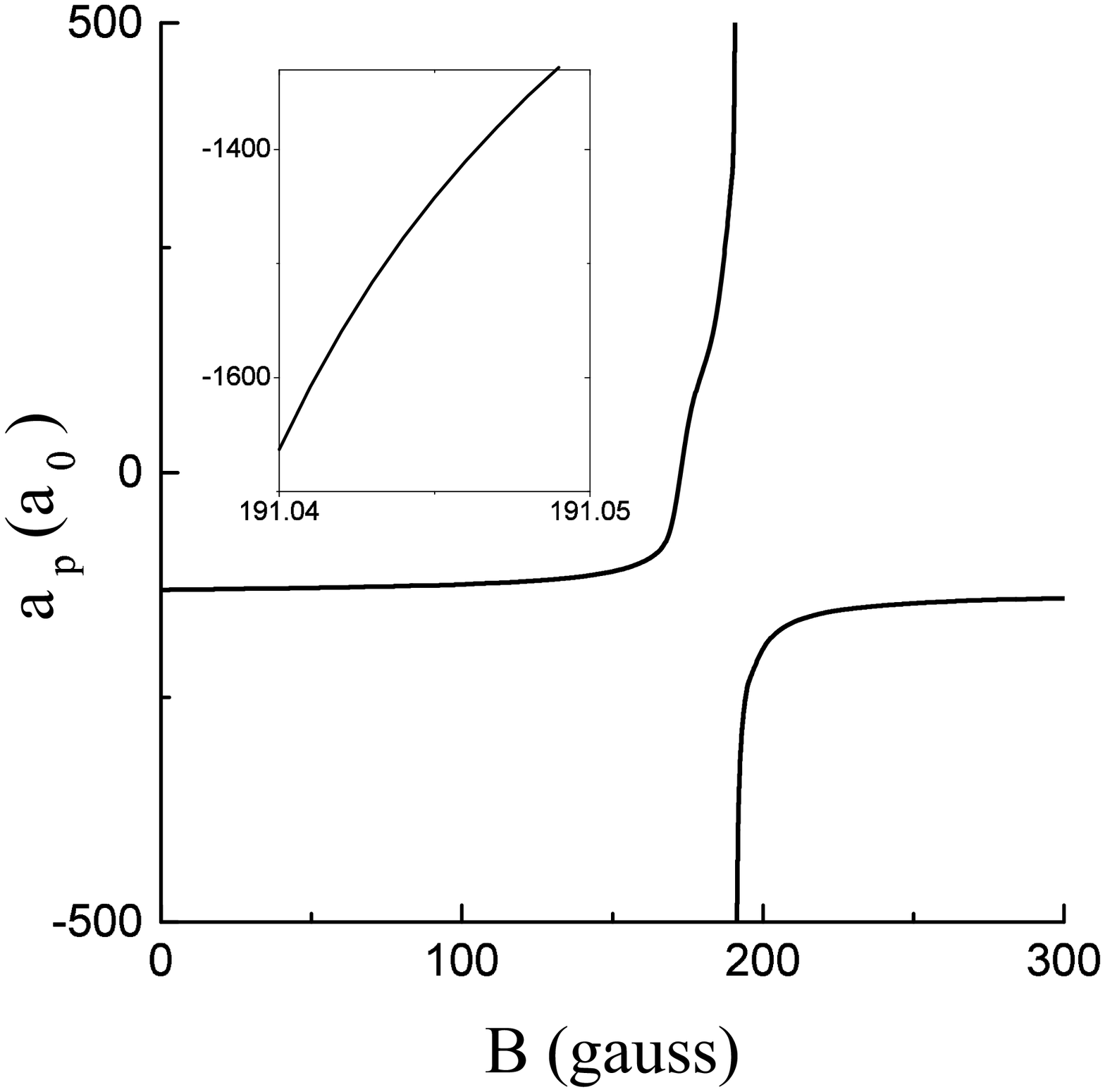}
\begin{figure}
\caption{Variation with magnetic field of the p-wave scattering length $a_p$
for $|9/2,-7/2 \rangle +$ $|9/2,-7/2 \rangle$ collisions of $^{40}$K.
Shown is the nominal case, where $a_p^{\rm triplet}(40) = -100$ $a_0$.
The inset focuses on the variation of $a_p$ with $B$ in the vicinity
of $a_p = -1500$ $a_0$.}
\end{figure}

\end{document}